\documentclass[aps,prb,twocolumn,amsmath,amssymb,groupedaddressx,longbibliography]{revtex4-2} 
\usepackage{here} \usepackage{graphicx} \usepackage{dcolumn} \usepackage{bm} \usepackage{multirow} \usepackage{amsmath} \usepackage{tabularx} \usepackage{amssymb} 
\usepackage[colorlinks=true,linkcolor=blue,citecolor=blue,urlcolor=blue]{hyperref} \usepackage{braket} \usepackage{color} \usepackage{ulem}  \usepackage{booktabs}

\begin{document} 
\title{Intrinsic anomalous Hall effect under \\ anisotropic magnetic dipole versus conventional magnetic dipole} 
\author{Satoru Ohgata}\author{Satoru Hayami} \affiliation{Graduate School of Science, Hokkaido University, Sapporo 060-0810, Japan.}
\begin{abstract} 
    We theoretically investigate the intrinsic anomalous Hall effect in two magnetically ordered systems: One is the ferromagnetic dipole system, and the other is the anisotropic magnetic dipole system, the latter of which has been proposed as a microscopic indicator of the anomalous Hall effect in antiferromagnets with the negligibly small magnetization. 
    We show their similarity and difference in the anomalous Hall effect by analyzing the fundamental tight-binding model on a two-dimensional square lattice. 
    We find that the magnitudes of the anomalous Hall effect in the two systems are similar to each other, while the magnetization in the anisotropic magnetic dipole system is much smaller than that in the ferromagnetic dipole system; this indicates that the microscopic mechanisms are different from each other. 
    We show that such a difference appears in the momentum-resolved Berry curvature resulting from the different types of magnetic order parameters. 
    We also show that the anomalous Hall effect in the anisotropic magnetic dipole system is enhanced when the magnitude of the spin-orbit coupling is smaller than that of the magnetic mean field. 
    Our results provide a possibility of the giant anomalous Hall effect in collinear and coplanar antiferromagnetic systems with the anisotropic magnetic dipole.
\end{abstract}

\maketitle
\section{Introduction}
    When an electric field is applied to a ferromagnetic (FM) material, a current flows perpendicular to the electric field and the magnetization. This phenomenon, known as the anomalous Hall effect (AHE), has been extensively studied in both theoretical and experimental contexts~\cite{Hall1879,Hall1881_AHE_discovery, Smit1958_skew_scattering_AHE, Maranzana_PhysRev.160.421, Berger_1970_side_jump_AHE, nozieres1973simple, Jungwirth_2002_AHE_FM_Semiconductors, Zeng_PhysRevLett.96.037204, Gosalbez_2015_Chiral_degen_Chern_num, Karplus1954_KL_intrinsic_AHE, Nagaosa_2010_AHE_review, Xiao_2010_BerryPhase}. Since the magnitude of the AHE is proportional to the magnetization of the material, it has traditionally been regarded as a phenomenon unique to FM materials. However, recent studies have demonstrated that it can also manifest in antiferromagnetic (AFM) materials, despite their nearly negligible magnetization; for instance, in collinear antiferromagnets (AFMs)~\cite{Solovyev_1997_MOKE, Sivadas_2016_MOKE, vsmejkal_2020Jun_CHE, Tenasini2020Apr_collinear_CoNbS, Yuan2020Jul_collinear_MnF2, Naka_2020Aug_AHE_kappa, Hayami2021AMD, smejkal2022altermag, Chen_PhysRevB.106.024421, Naka_PhysRevB.106.195149, Gonzalez_2023_collinear_alpha_MnTe, Attias_2024_AHE_Altermagnet, Reimers2024_collinear_CrSb, Sato_2024_AHE_Altermagnet, Kurita_2024_XMCD_AFM} and noncollinear AFMs~\cite{Ohgushi_2000_QHE_kagome, Shindou_2001_AHE_orbital_Ferromagnetism, Tomizawa_2009_AHE_kagome_orbital_ABE, Chen_2014_AHE_noncollinear, Nakatsuji2015_non_collinear_Mn3Sn, Nakatsuji2016_non_collinear_Mn3Ge, Suzuki_2017_AHE_cluster_multipole, Higo2018_non_collinear_Mn3Sn,  Chen_2020_AHE_AFM_m_field_control}. These types of AFMs have been attracting significant interest in the field of spintronics, since they do not produce a stray field and exhibit a fast switching speed of less than 10 picoseconds, which is 100 times faster than in ferromagnets (FMs)~\cite{miwa2021_MOKE_switching_speed_measurement}. These features make them promising candidate materials for next-generation spintronic devices, offering high-density integration and ultra-fast performance~\cite{Chen2024AFM_spintronics}.

\begin{figure}[t] 
        \centering \includegraphics[width=0.45\textwidth]{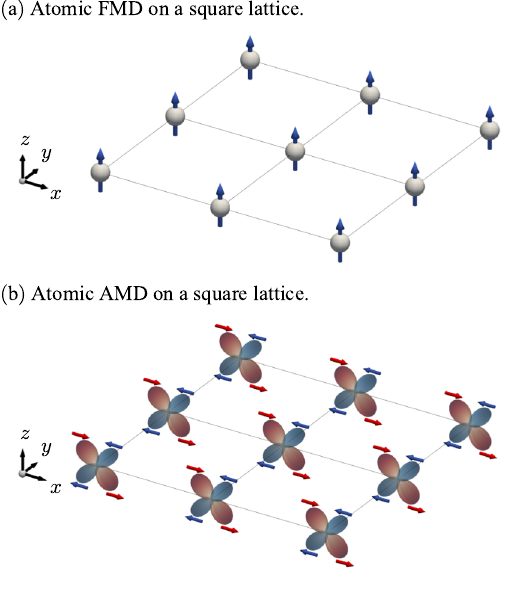}       
        \caption{
            (a) Schematic picture of ferromagnetic dipole (FMD) ordering, where the spins $s_z$ are ordered at each site. 
            (b) Schematic picture of anisotropic magnetic dipole (AMD) ordering, which consists of the product of the orbital $Q_{zx}$ and the spin $s_x$, i.e., $M_{z}'^{(s_x)}$. 
            This ordering induces the anomalous Hall conductivity $\sigma_{xy}$, just as the FMD ordering shown in panel (a) does. 
            These figures were drawn by QtDraw~\cite{Kusunose_2023_symmetry_adopted_modeling}.
        }\label{fig:FMD_AMD}
    \end{figure}

    From a microscopic viewpoint, key electronic degrees of freedom to induce the AHE in AFMs differ from those in FMs. Based on the symmetry-adapted multipole representation, which describes physical quantities using a complete basis set of four types of multipoles~\cite{kusunose2022generalization, hayami2024unified}, it was demonstrated that the anisotropic magnetic dipole (AMD) is the key indicator for the AHE in AFMs~\cite{Hayami2021AMD}. The AMD is a time-reversal-odd rank-1 axial vector belonging to the same irreducible representation as conventional magnetic dipoles such as spin and orbital angular momenta, and is related to the $\bm{T}$-vector in the context of x-ray magneto-circular dichroism~\cite{Carra1993_AMD_in_XMCD, Stohr1995a_AMD_in_XMCD, Stohr1995b_AMD_in_XMCD, Crocombette1996_AMD_in_XMCD, Yamasaki2020_AMD, Sasabe_2021_XMCD_AFM, Sasabe_2024_AMD_t2g}. The operator expression of the AMD in an atomic scale is given as follows~\cite{kusunose2020complete}: 
    \begin{align} \label{eq:AMD} \hat{\bm{M}}' = \frac{1}{\sqrt{10}} \left[ 3(\hat{\bm{r}} \cdot \hat{\bm{\sigma}})\hat{\bm{r}} - \hat{r}^{2}\hat{\bm{\sigma}} \right] , \end{align} where $\hat{\bm{\sigma}}$ is the vector composed of the Pauli matrices, which act on spin space and $\hat{\bm{r}}$ is the position operator. The AMD is decomposed as: 
    \begin{align} \label{eq:AMD_explicit_form} 
        \begin{bmatrix} \hat{M}_x' \\ \hat{M}_y' \\ \hat{M}_z' \end{bmatrix} = \sqrt{\frac{3}{10}} 
        \begin{bmatrix} \frac{-1}{\sqrt{3}}\hat{Q}_u + \hat{Q}_v & \hat{Q}_{xy} & \hat{Q}_{zx} \\ \hat{Q}_{xy} & \frac{-1}{\sqrt{3}}\hat{Q}_u + \hat{Q}_v & \hat{Q}_{yz} \\ \hat{Q}_{zx} & \hat{Q}_{yz} & \frac{2}{\sqrt{3}} \hat{Q}_u \end{bmatrix} 
        \begin{bmatrix} \hat{\sigma}_x \\ \hat{\sigma}_y \\ \hat{\sigma}_z \end{bmatrix},
    \end{align} where $\hat{Q}_u, \hat{Q}_v, \hat{Q}_{yz}, \hat{Q}_{zx}, \hat{Q}_{xy}$ are the electric quadrupole operators with the functional forms of $(3\hat{z}^2 - \hat{r}^2)/2, \sqrt{3}(\hat{x}^2 - \hat{y}^2)/2, \sqrt{3}\hat{y}\hat{z}, \sqrt{3}\hat{z}\hat{x}, \sqrt{3}\hat{x}\hat{y}$, respectively. Thus, the AMD is linked to a quadrupole distribution of spins, which does not exhibit net magnetization. This expression indicates that the AHE can even occur when the spin moments lie within a plane defined by the input electric field and the output current, provided the AMD has a perpendicular component. For instance, the $zx$($yz$)-type spatial distribution of the spin component $s_x$ ($s_y$) results in an anomalous Hall conductivity tensor of $\sigma_{xy} = -\sigma_{yx}$ owing to a finite AMD in the $z$ direction, where the magnetic symmetry is the same as that in the FM state with the $z$-spin polarization. Figure~\ref{fig:FMD_AMD} provides a schematic picture of this scenario. The concept of the AMD can be applied from atomic scale ~\cite{kusunose2020complete} to cluster scale~\cite{Hayami2021AMD}.

    In the present study, we theoretically investigate the role of the atomic-scale AMD in the AHE, emphasizing the similarities and differences compared to systems with FM dipoles (FMD), i.e., the conventional magnetic dipoles. We analyze the anomalous Hall conductivity of a fundamental multi-orbital model that includes both the FMD and AMD degrees of freedom on a square lattice based on the linear response theory. Our findings show that the AMD can achieve a large AHE comparable to that under the FMD, with a much smaller magnetization. To understand this difference, we examine the momentum-resolved Berry curvature in each case and identify distinct trends associated with gap opening. Additionally, we demonstrate that the essential model parameters for the magnetization under the AMD differ from those under the FMD. Furthermore, we show that the AHE under the AMD tends to be enhanced when the spin-orbit coupling (SOC) is weaker than that of the magnetic mean field. Our results provide insights into realizing AFMs with a giant AHE.

    The rest of this paper is organized as follows. In Sec.~\ref{sec:method}, we introduce the model Hamiltonian and lattice structure, along with the calculation methods for the magnetization, the Berry curvature, and the AHE. In Sec.~\ref{subsec:AHE_origin}, we compare the calculation results of the AHE and magnetization under the FMD and AMD. By calculating the densities of Berry curvature in each case, we demonstrate that the source of the AHE in the AMD-ordered system differs from that in the FMD-ordered system. Then, in Sec.~\ref{subsec:Parameter_dependence}, we present the dependence of the AHE and magnetization on the SOC constant and magnetic mean field. Here, we show that the peak values of the AHE in the two systems are similar, while the magnetization in the AMD-ordered system is much smaller than that in the FMD-ordered system. Additionally, we highlight that the AHE in the AMD-ordered system is enhanced when the magnitude of the SOC is smaller than that of the magnetic mean field. Section~\ref{sec:summary} is devoted to the summary of the present paper.

\section{Model and method}\label{sec:method}
    We construct a tight-binding model to investigate the differences in the AHE under the FMD and AMD. To this end, we employ a minimal multi-orbital model composed of three $p$-orbitals $(p_x, p_y, p_z)$ with spin $(\uparrow, \downarrow)$ on a two-dimensional square lattice under the point group $D_{\rm 4h}$, which incorporates both the FMD and AMD degrees of freedom within a low-energy physical space. The tight-binding Hamiltonian is expressed as follows: 
    \begin{align} \label{eq:Hamiltonian} 
        \mathcal{\hat{H}} &= \mathcal{\hat{H}}_t + \mathcal{\hat{H}}_{\rm SOC} + \mathcal{\hat{H}}_{\rm FMD} + \mathcal{\hat{H}}_{\rm AMD}, \\ 
        \mathcal{\hat{H}}_t&= \sum_{\bm{k}\alpha\alpha'\sigma} \varepsilon_{\bm{k}\alpha\alpha'} \hat{c}^{\dagger}_{\bm{k}\alpha\sigma}\hat{c}_{\bm{k}\alpha'\sigma}, \\ 
        \mathcal{\hat{H}}_{\rm SOC} &=\lambda \sum_{\bm{k}\alpha\alpha'\sigma\sigma'} (\hat{\bm{l}} \cdot \hat{\bm{s}})^{\sigma\sigma'}_{\alpha\alpha'} \hat{c}^{\dagger}_{\bm{k}\alpha\sigma}\hat{c}_{\bm{k}\alpha'\sigma'}, \\ 
        \mathcal{\hat{H}}_{\rm FMD} &= -h_{\rm FMD} \sum_{\bm{k}\alpha\sigma}(\hat{s}_z)_{\alpha\alpha}^{\sigma\sigma} \hat{c}^{\dagger}_{\bm{k}\alpha\sigma}\hat{c}_{\bm{k}\alpha\sigma}, \\ 
        \mathcal{\hat{H}}_{\rm AMD} &= -h_{\rm AMD} \sum_{\bm{k}\alpha\alpha'\sigma\sigma'}(\hat{M}_{z}'^{(s_x)})_{\alpha\alpha'}^{\sigma\sigma'} \hat{c}^{\dagger}_{\bm{k}\alpha\sigma}\hat{c}_{\bm{k}\alpha'\sigma'}, 
    \end{align} where $\hat{c}^{\dagger}_{\bm{k}\alpha\sigma}$ ($\hat{c}_{\bm{k}\alpha\sigma}$) is the creation (annihilation) operator for electrons with wave vector $\bm{k}$, $p_{\alpha}$-orbital ($\alpha = x, y, z$), and spin $\sigma = \uparrow, \downarrow$. The first term, $\mathcal{\hat{H}}_t$, represents the kinetic Hamiltonian, which includes the Slater-Koster parameters $t_{pp\sigma}$ and $t_{pp\pi}$ for the nearest-neighbor hopping and $t'_{pp\sigma}$ and $t'_{pp\pi}$ for the next-nearest-neighbor hopping. Specifically, $\varepsilon_{\bm{k}\alpha\alpha'}$ is related to these hopping parameters as follows:
    \begin{align} \label{eq:kinetic Hamiltonian} 
        \varepsilon_{\bm{k}xx} &= 2 t_{pp\sigma} \cos k_x + 2 t_{pp\pi} \cos k_y \\ &\quad + 2 (t'_{pp\sigma} +t'_{pp\pi}) \cos k_x \cos k_y , \notag \\
        \varepsilon_{\bm{k}yy} &= 2 t_{pp\sigma} \cos k_y + 2 t_{pp\pi} \cos k_x \\ &\quad + 2 (t'_{pp\sigma} +t'_{pp\pi}) \cos k_x \cos k_y , \notag \\ 
        \varepsilon_{\bm{k}zz} &= 2 t_{pp\pi} (\cos k_x + \cos k_y) + 4 t'_{pp\pi} \cos k_x \cos k_y , \\ 
        \varepsilon_{\bm{k}xy} &= -2 (t'_{pp\sigma} - t'_{pp\pi}) \sin k_x \sin k_y , \\ 
        \varepsilon_{\bm{k}yx} &= \varepsilon_{\bm{k}xy} , \\ 
        \varepsilon_{\bm{k}\alpha\alpha'} &= 0 \quad (\text{otherwise}), 
    \end{align} where we set the lattice constant to unity. Note that these satisfy the fourfold rotational symmetry of the lattice structure. In the following, the Slater-Koster parameters are fixed to $(t_{pp\sigma}, t_{pp\pi}, t'_{pp\sigma}, t'_{pp\pi}) = (1, -0.5, 0.6, -0.3)$. The second term, $\mathcal{\hat{H}}_{\rm SOC}$, represents the relativistic SOC with the coupling constant $\lambda$; $\hat{\bm{s}}=2\hat{\bm{\sigma}}$ and $\hat{\bm{l}}$ represent the dimensionless spin and orbital angular momentum operators, respectively.

    The third and fourth terms, $\mathcal{\hat{H}}_{\rm FMD}$ and $\mathcal{\hat{H}}_{\rm AMD}$, describe the mean-field Hamiltonian for the FMD and AMD along the $z$ direction, with $h_{\rm FMD}$ and $h_{\rm AMD}$ representing the magnitude of each mean field. From the symmetry viewpoint, the FMD and AMD belong to the same irreducible representation under the point group $D_{\rm 4h}$. For the AMD, we consider the $x$-spin contribution to $\hat{M}'_z$ in Eq.~(\ref{eq:AMD_explicit_form}), specifically $\hat{Q}_{zx} \hat{s}_x$, with the collinear spin configuration implicitly in mind; we use the notation $\hat{M}_{z}'^{(s_x)} \equiv \hat{Q}_{zx} \hat{s}_x$ here and hereafter. It is noteworthy that the AMD is regarded as the spin-orbital entangled quantity. The schematic pictures of the FMD and AMD are shown in Figs.~\ref{fig:FMD_AMD}(a) and \ref{fig:FMD_AMD}(b), respectively.

\begin{figure}[t] 
        \centering \includegraphics[width=0.50\textwidth]{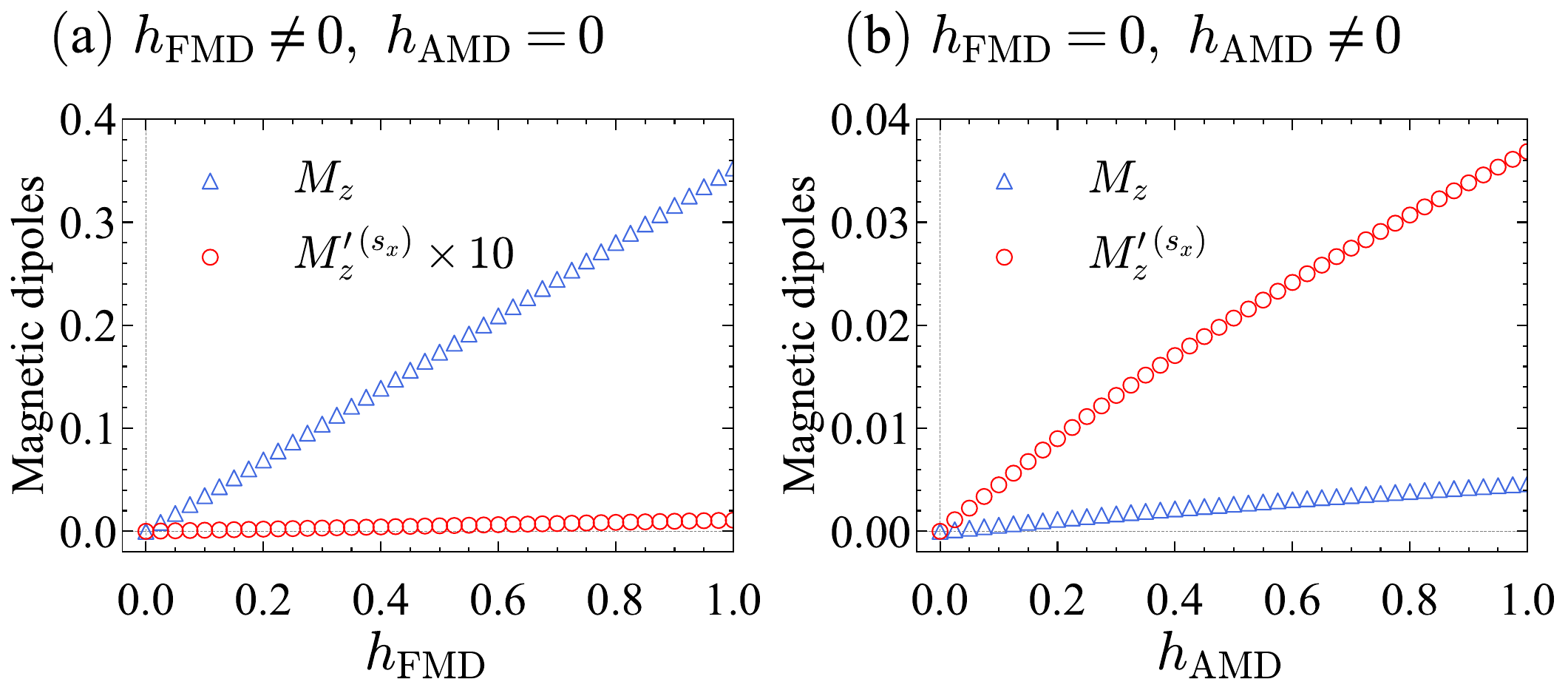} 
        \caption{(a, b) Expectation values of the FMD and AMD, $M_z$ and $M_{z}'^{(s_x)}$, as functions of the magnetic mean field of (a) the FMD $h_{\rm FMD}$ and (b) the AMD $h_{\rm AMD}$. The SOC constant and chemical potential are set to $\lambda=0.05$ and $\mu=-1$, respectively.}
        \label{fig:Orthognarity}
    \end{figure}

    In order to understand the relationship between the FMD and AMD, we show the expectation values of their ordered parameters when either $h_{\rm FMD}$ or $h_{\rm AMD}$ is nonzero in Figs.~\ref{fig:Orthognarity}(a) and \ref{fig:Orthognarity}(b). We set the primary order parameter for the FMD as $M_z$, which is given by 
    \begin{align} \label{eq:M_z} M_z &= \frac{1}{N_{\bm{k}}} \sum_{\bm{k}n} f(\mu, E_{n\bm{k}}) M_{z,n\bm{k}} , \end{align} where  $N_{\bm{k}}$ is the number of the grid points in the Brillouin zone, $f(\mu, E_{n\bm{k}}) = 1/\left[ \exp\left\{ \beta (E_{n\bm{k}} - \mu) \right\} + 1 \right]$ represents the Fermi-Dirac distribution function with the chemical potential $\mu$, the $n$th-band eigenvalue $E_{n\bm{k}}$, the inverse temperature $\beta=1/T$ (the Boltzmann constant is set to unity), and
    \begin{align} \label{eq:M_znk} M_{z,n\bm{k}} &= \langle \psi_{n\bm{k}} | \mu_{\rm B} (\hat{l}_z + 2\hat{s}_z) | \psi_{n\bm{k}} \rangle, \end{align} where $\psi_{n\bm{k}}$ is the eigenstate, and $\hat{s}_z = \sigma_z / 2$ and $\hat{l}_z$ represent the $z$-components of spin and orbital angular momentum operators, respectively. $M_z$ corresponds to the total magnetization of the system.

    Meanwhile, the primary order parameter for the AMD corresponds to $\hat{M}_{z}'^{(s_x)}=\hat{Q}_{zx} \hat{s}_x$, as introduced above. The calculation method for the expectation value of this operator is the same as that used in Eq.~(\ref{eq:M_z}), i.e.,
    \begin{align} \label{eq:AMD_thermal_avg} M_{z}'^{(s_x)} &= \frac{1}{N_{\bm{k}}} \sum_{\bm{k}n} f(\mu, E_{n\bm{k}}) M_{n\bm{k},z}'^{(s_x)}, \end{align} where 
    \begin{align} \label{eq:AMD_znk} M_{n\bm{k},z}'^{(s_x)} &= \langle \psi_{n\bm{k}} | \hat{Q}_{zx} \hat{s}_x | \psi_{n\bm{k}} \rangle. \end{align} We set $\mu_{\rm B}=1$ and $\hbar=1$ in the following calculations.

    As shown in Fig.~\ref{fig:Orthognarity}(a), $M_{z}'^{(s_x)} \ll M_z$ holds under the FMD ordering, where $h_{\rm FMD} \neq 0$ and $h_{\rm AMD}=0$. This indicates that $M_{z}'^{(s_x)}$ is the secondary order parameter under the FMD ordering. The small but nonzero values of $M_{z}'^{(s_x)}$ are owing to the presence of the SOC, which couples them at the microscopic level. The opposite behavior is found in the AMD ordering [Fig.~\ref{fig:Orthognarity}(b)] for $h_{\rm FMD} = 0$ and $h_{\rm AMD} \neq 0$. It is noteworthy that these secondary order parameters, either $M_z$ under the AMD phase or $M_{z}'^{(s_x)}$ under the FMD phase, become zero when the $x$ component of the SOC, i.e., $\hat{l}_x \hat{s}_x$, is neglected [see Eq.~(\ref{eq:Mz_under_AMD})].

    In order to discuss the similarity and difference of the AHE between the FMD and AMD orderings, we calculate the anomalous Hall conductivity based on the linear response theory, which is given by~\cite{Nagaosa_2010_AHE_review}
    \begin{align} \label{eq:sigma_xy} \sigma_{xy} &= - \frac{e^2}{\hbar} \frac{1}{N_{\bm{k}}} \sum_{n\bm{k}} f(\mu, E_{n\bm{k}}) \Omega_{xy,n\bm{k}} , \end{align} where $e$ is the elementary charge and $\Omega_{xy,n\bm{k}}$ is the Berry curvature given by 
    \begin{align} \label{eq:Omega_znk} \Omega_{xy,n\bm{k}} &= -2 \hbar^2 \, \sum_{m \neq n} \frac{ \text{Im} \left[\langle \psi_{m\bm{k}} | \hat{v}_{x,\bm{k}} | \psi_{n\bm{k}} \rangle \langle \psi_{n\bm{k}} | \hat{v}_{y,\bm{k}} | \psi_{m\bm{k}} \rangle \right] } { (E_{n\bm{k}} - E_{m\bm{k}})^2 + (\hbar/2\tau)^2 }. \end{align} $\hat{\bm{v}} \equiv \partial\mathcal{\hat{H}}/\partial (\hbar \bm{k})$ denotes the velocity operator and $\tau^{-1}$ is the scattering rate; we adopt the relaxation time approximation neglecting the side-jump and skew-scattering contributions~\cite{Onoda2008AHE_int_ext_cross_over}. Additionally, to identify the eigenstates giving the dominant contribution to $\sigma_{xy}$, we calculate the momentum-resolved Berry curvature at each $\bm{k}$ in the Brillouin zone:
    \begin{align} \label{eq:Berry curvature expectation} \Omega_{xy}(\bm{k}) &= \sum_{n} f(\mu, E_{n\bm{k}}) \Omega_{xy,n\bm{k}} , \end{align} and the density of the positive and negative Berry curvature at a energy level $E$:
    \begin{align} \label{eq:Berry curvature density} \rho[\Omega_{xy}^{\pm}] &= \frac{1}{N_{\bm{k}}} \sum_{n\bm{k}} \delta(E - E_{n\bm{k}}) \theta(\pm \Omega_{xy,n\bm{k}}) \Omega_{xy,n\bm{k}}, \end{align} where $\delta$ and $\theta$ represent the Dirac delta function and the Heaviside step function, respectively. $e$ is set to 1, $N_{\bm{k}}$ is set to $400^2$, $(\hbar/2\tau)$ is set to $10^{-3}$, and $T$ is set to $10^{-2}$. We especially focus on the behavior of $\sigma_{xy}$ while varying the magnetic mean fields $h_{\rm FMD}$, $h_{\rm AMD}$ and the SOC constant $\lambda$.

    \begin{figure*}[ht] 
        \centering \includegraphics[width=\textwidth]{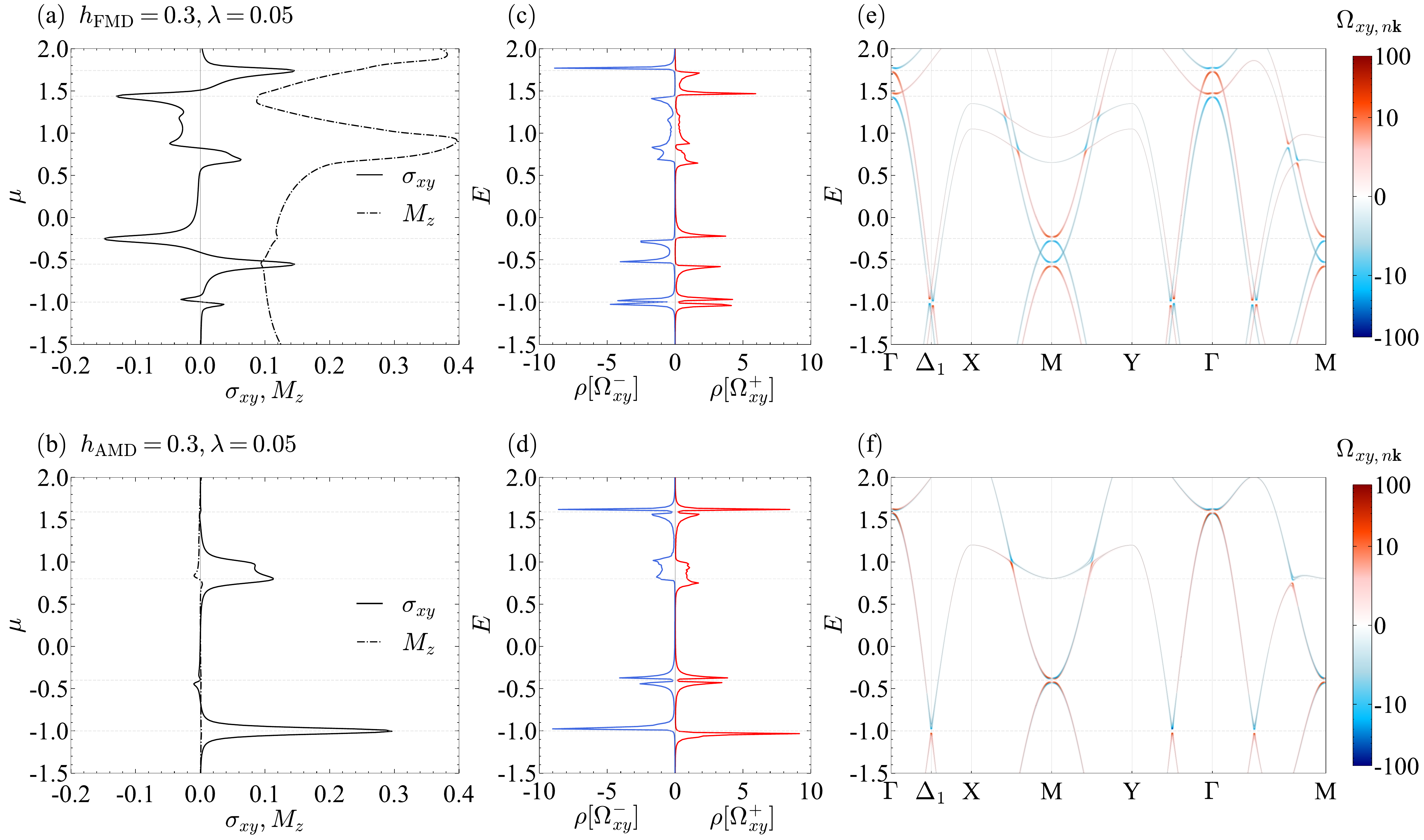}        
        \caption{
            (a, b) Anomalous Hall conductivity $\sigma_{xy}$ and total magnetization $M_z$ as functions of the chemical potential $\mu$.     
            (c, d) Density of Berry curvature $\rho[\Omega_{xy}^{\pm}]$ as a function of Energy $E$. 
            (e, f) Energy band mapped with the Berry curvature $\Omega_{xy,n\bm{k}}$; see the Brillouin zone in Fig.~\ref{fig:Enlarged_view}(a) for the notation of the horizontal axis.
            (a, c, e) are calculated under the FMD ordering with magnetic mean fields $h_{\rm FMD} = 0.3$ and $h_{\rm AMD} = 0$, while (b, d, e) are calculated under the AMD ordering with $h_{\rm FMD} = 0$ and $h_{\rm AMD} = 0.3$.
            The SOC constant is set to $\lambda = 0.05$ for (a--f). 
        }\label{fig:Hall_conductivity_and_its_source}
    \end{figure*}

    \begin{figure*}[ht] 
        \centering \includegraphics[width=\textwidth]{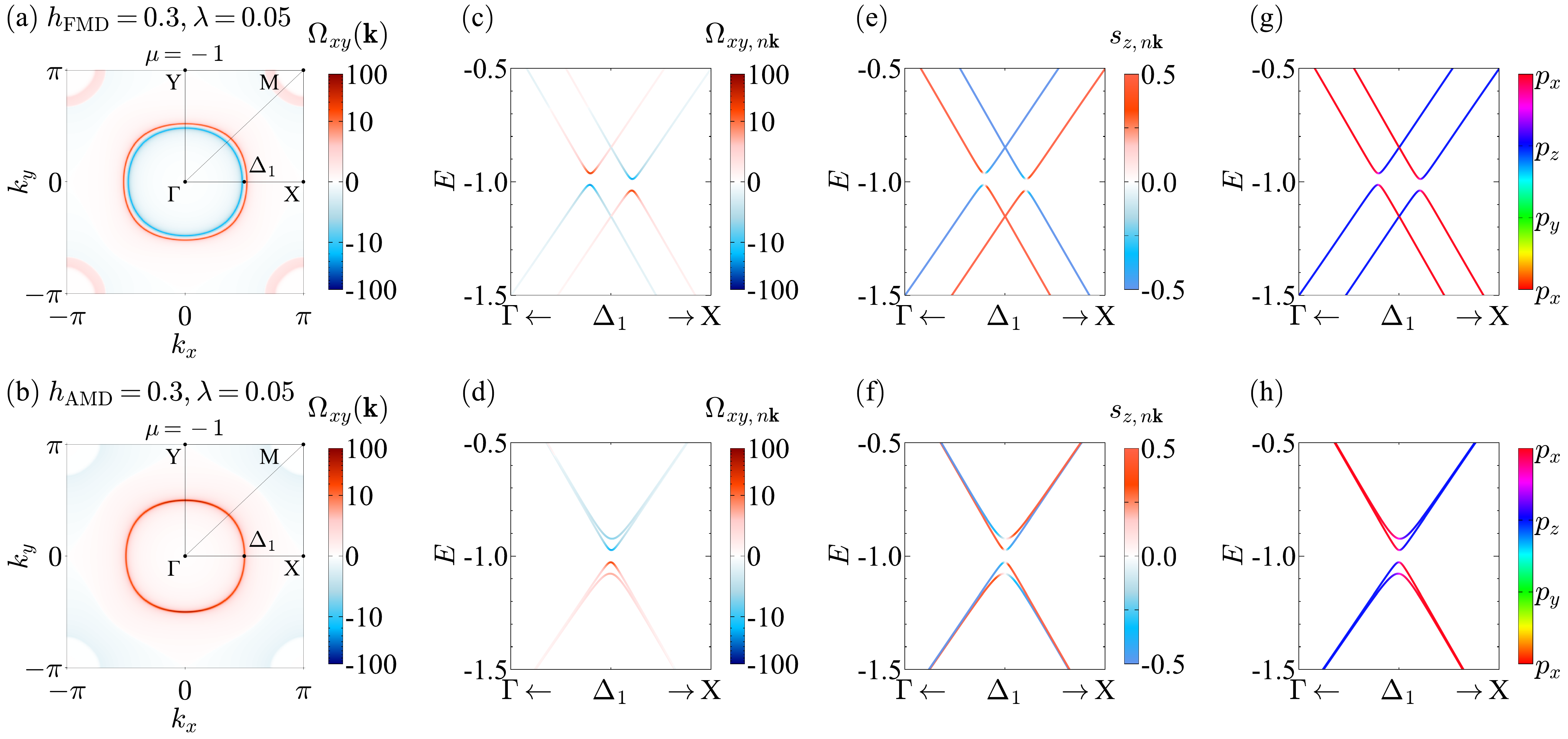}        
        \caption{
            (a, b) Expectation values of the Berry curvature $\Omega_{xy}(\bm{k})$ at chemical potential $\mu = -1$, mapped in the Brillouin zone.
            (c--h) Energy bands around the $\Delta_1$ point mapped with the following quantities:
            (c, d) Berry curvature $\Omega_{xy,n\bm{k}}$, 
            (e, f) $z$-component of spin $s_{z,n\bm{k}} \equiv \langle \psi_{n\bm{k}} | \hat{s}_z | \psi_{n\bm{k}} \rangle$, and 
            (g, h) orbital component $\sum_{\sigma}|\langle p_{\alpha\sigma} | \psi_{n\bm{k}} \rangle|^{2}$.
            The SOC constant is set to $\lambda = 0.05$ for (a--f).
            The magnetic mean fields are set to $h_{\rm FMD} = 0.3, h_{\rm AMD} = 0$ for (a, c, e, g), and $h_{\rm FMD} = 0, h_{\rm AMD} = 0.3$ for (b, d, f, h).
        }\label{fig:Enlarged_view}
    \end{figure*}

\section{Results}\label{sec:results}
    In Sec.~\ref{subsec:AHE_origin}, we compare the magnitude of $\sigma_{xy}$ and $M_z$ under the FMD and AMD orderings. We also discuss the origin of $\sigma_{xy}$ by analyzing the momentum-resolved Berry curvature $\Omega_{xy,n\bm{k}}$ and its sign-resolved density $\rho[\Omega_{xy}^{\pm}]$. In Sec.~\ref{subsec:Parameter_dependence}, we present the dependence of $\sigma_{xy}$ and $M_z$ on the SOC constant $\lambda$ and the magnetic mean fields $h_{\rm FMD}, h_{\rm AMD}$.

\subsection{Origin of the AHE}\label{subsec:AHE_origin} 
    First, we analyze the behavior of $\sigma_{xy}$ and $M_z$ as functions of $\mu$, and compare their magnitudes between the FMD and AMD orderings. Figure~\ref{fig:Hall_conductivity_and_its_source}(a) shows $\sigma_{xy}$ and $M_z$ calculated for the FMD-ordered system at $h_{\rm FMD} = 0.3$ and $\lambda = 0.05$. For almost all $\mu$, $\sigma_{xy}$ and $M_z$ exhibits nonzero values except for the $\mu$ values at which $\sigma_{xy}$ changes its sign. The maximum value of $|\sigma_{xy}|$ is around 0.15 at $\mu = -0.25$, with $M_z = 0.12$.

    On the other hand, a distinct feature appears in the AMD-ordered system; the total magnetization $M_z$ almost vanishes over the entire range of $\mu$ under the AMD ordering, as shown in Fig.~\ref{fig:Hall_conductivity_and_its_source}(b). Nevertheless, $\sigma_{xy}$ under the AMD ordering can reach values comparable to those under the FMD ordering. For instance, the maximum value of $|\sigma_{xy}|$ is around 0.30 at $\mu = -1.0$ despite the small magnetization $M_z = 0.0017$. From these results, it can be inferred that the AMD serves as a source of the giant AHE with a negligibly small $|M_z|$.

    In order to understand the distinct behavior of $\sigma_{xy}$ between the FMD and AMD orderings, we examine the Berry curvature in each case. Figure~\ref{fig:Hall_conductivity_and_its_source}(c) shows the density of Berry curvature in the FMD-ordered system corresponding to Fig.~\ref{fig:Hall_conductivity_and_its_source}(a), where we decompose the density of Berry curvature into the positive and negative parts $\rho[\Omega_{xy}^{\pm}]$ at each energy (chemical potential); a significant rate of change in $\sigma_{xy}$ with respect to $\mu$ is expected when the difference between the positive and negative densities of Berry curvature ($\rho[\Omega_{xy}^{+}]-|\rho[\Omega_{xy}^{-}]|$) is large. We also plot the corresponding band structure mapped with the momentum-resolved Berry curvature $\Omega_{xy,n\bm{k}}$ in Fig.~\ref{fig:Hall_conductivity_and_its_source}(e) in order to identify which eigenstates give dominant contributions to $\sigma_{xy}$. One finds that the peaks of $|\sigma_{xy}|$ originate from the M point at $E = -0.25, -0.55$ and the $\Gamma$ point at $E = 1.44, 1.74$, where the small gaps of $0.05(=\lambda)$ opens due to the SOC. The large enhancement of Berry curvature in the small-gap region arises from the denominator in Eq.~(\ref{eq:Omega_znk}), which depends on the energy difference. Meanwhile, it is noted that the small band gap does not always lead to large $\sigma_{xy}$. For example, $\sigma_{xy}$ is suppressed at $\mu = -1$, despite the presence of the band anticrossing at $\Delta_1$ along the $\Gamma$-X line. This is because the opposite signs of Berry curvature are generated at almost the same energy level close to the band gap, as shown in Fig.~\ref{fig:Hall_conductivity_and_its_source}(e) around $\Delta_1$ and $E = -1$; see also Fig.~\ref{fig:Enlarged_view}(c) for an enlarged view.

    To understand the mechanism of gap opening and its relation to the sign of Berry curvature in Fig.~\ref{fig:Enlarged_view}(c), we consider the effective four-band model around $\Delta_1$ and $E = -1$~\cite{Lei2023_Nodal_ring}. Starting from $h_{\rm FMD} = \lambda = 0$, a four-fold degeneracy occurs at $\Delta_1$, which is a crossing point of $p_{x\uparrow\downarrow}$ bands and $p_{z\uparrow\downarrow}$ bands. When $h_{\rm FMD}$ is introduced, both $p_{x\uparrow\downarrow}$ bands and $p_{z\uparrow\downarrow}$ bands split by $\pm h_{\rm FMD}/2$ globally, in accordance with the mean-field Hamiltonian of the FMD, which can be written in an explicit form: 
    \begin{align} \label{eq:FMD matrix elements} \mathcal{\hat{H}}_{\rm FMD} = -\frac{h_{\rm FMD}}{2} \sum_{\alpha=x,y,z} \left( \hat{c}^{\dagger}_{  \alpha\uparrow} \hat{c}_{  \alpha\uparrow} - \hat{c}^{\dagger}_{  \alpha\downarrow} \hat{c}_{  \alpha\downarrow}\right) . \end{align} Thus, a rhombus shape is created around $\Delta_1$. Then, when the $y$ component of the SOC: 
    \begin{align} \label{eq:lysy matrix elements} \lambda \hat{l}_y \hat{s}_y &= \frac{\lambda}{2}    \left( \hat{c}^{\dagger}_{  z\uparrow} \hat{c}_{  x\downarrow} - \hat{c}^{\dagger}_{  z\downarrow} \hat{c}_{  x\uparrow} \right) + \text{h.c.}, \end{align} is further introduced, the degeneracy is lifted by $\pm\lambda$/2 from the crossing point between the $p_{x\downarrow}$-$p_{z\uparrow}$ bands and $p_{x\uparrow}$-$p_{z\downarrow}$ bands, forming a partially gapped rhombus hybridizing the spin and orbital components, as shown in Figs.~\ref{fig:Enlarged_view}(e) and \ref{fig:Enlarged_view}(g). As shown in Fig.~\ref{fig:Enlarged_view}(c), these gaps generate opposite signs of Berry curvature because the $\hat{c}^{\dagger}_{  z\downarrow} \hat{c}_{  x\uparrow}$ and $\hat{c}^{\dagger}_{  z\uparrow} \hat{c}_{  x\downarrow}$ components of $\lambda \hat{l}_y \hat{s}_y$ have opposite effects on each crossing. Since such gaps surround the $\Gamma$ point, rings with positive and negative Berry curvature are formed when $\mu$ is set within these gaps, as shown in Fig.~\ref{fig:Enlarged_view}(a). These rings cancel out each other, resulting in the almost vanishing Hall conductivity, i.e., $\sigma_{xy} \approx 0$ , at $\mu = -1$ [Fig.~\ref{fig:Hall_conductivity_and_its_source}(a)].

    \begin{figure*}[t]
        \centering \includegraphics[width=0.85\textwidth]{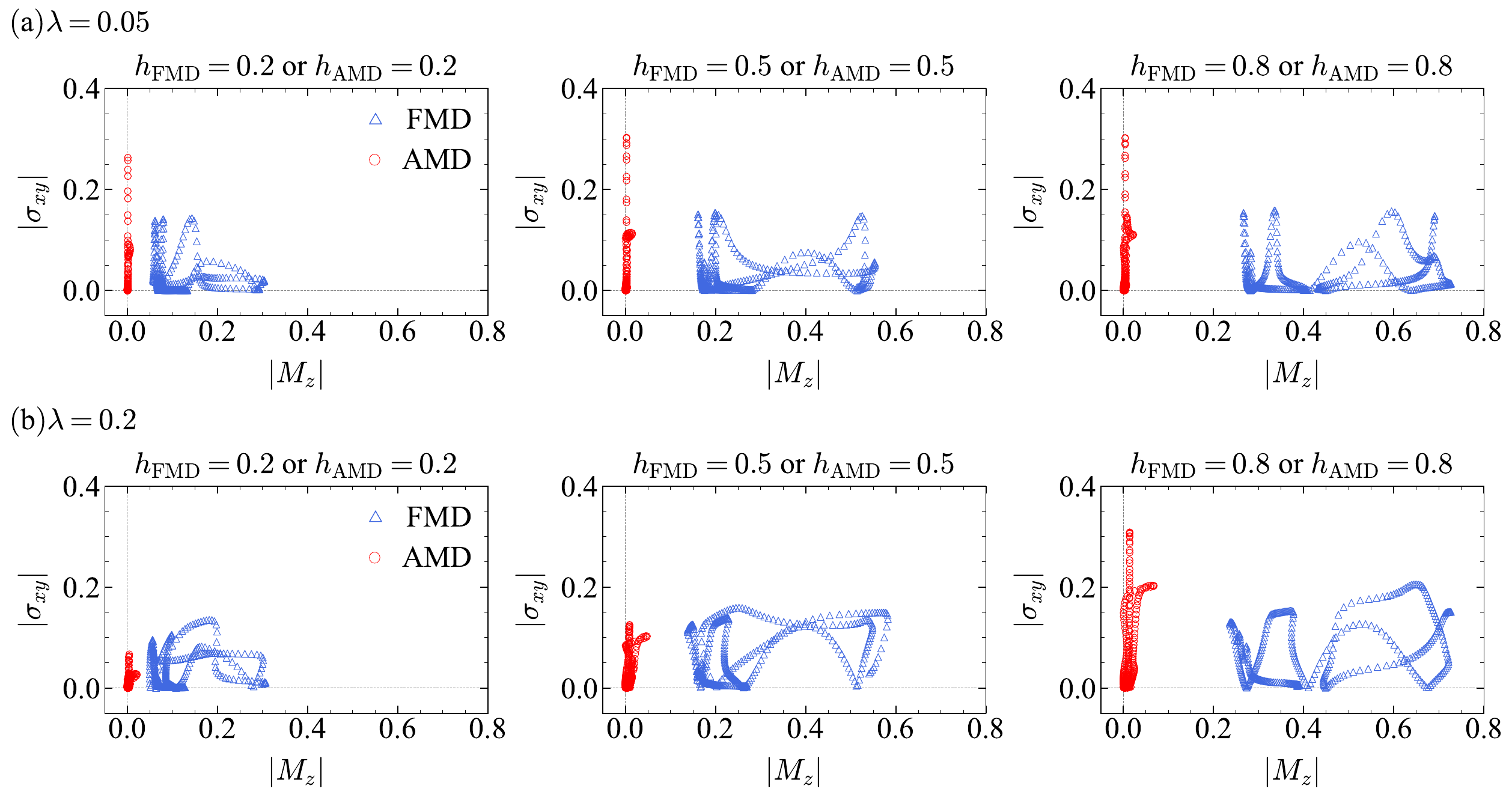}    
        \caption{
            The anomalous Hall conductivity $|\sigma_{xy}|$ and the total magnetization $|M_z|$ calculated for various values of magnetic mean fields: either $h_{\rm FMD}=0.2, 0.5, 0.8$ or $h_{\rm AMD}=0.2, 0.5, 0.8$.
            Plots for the FMD orderings are shown as blue triangles, while the AMD orderings are shown as red circles. 
            The SOC constant is set to $\lambda = 0.05$ for (a) and $\lambda = 0.2$ for (b). 
            The data is plotted for varying values of the chemical potential $\mu$, ranging from $-2$ to $2$.
        }\label{fig:Mz_sigma_plot}
    \end{figure*}

    Next, we discuss the behavior of $\sigma_{xy}$ under the AMD. In contrast to the FMD case, $\sigma_{xy}$ under the AMD is enhanced in the vicinity of the band anticrossing at low-symmetry points in the Brillouin zone rather than high-symmetry points. For example, $\sigma_{xy}$ is largely enhanced around $\mu=-1$ [Fig.~\ref{fig:Hall_conductivity_and_its_source}(b)], which arises from the band anticrossings including the $\Delta_1$ point [Fig.~\ref{fig:Hall_conductivity_and_its_source}(f)]. This is clearly illustrated in Fig.~\ref{fig:Enlarged_view}(d), where a pair of small gaps with a large Berry curvature emerges at $\Delta_1$. Since such gaps surround the $\Gamma$ point, a ring with positive Berry curvature is formed when the $\mu$ is set within the gap, as shown in Fig.~\ref{fig:Enlarged_view}(b). This ring makes the dominant contribution to $\sigma_{xy}$ at $\mu=-1$.

    The distinct behavior of Berry curvature around the gap opening from the FMD is attributed to the difference in the matrix elements of the order parameters for the FMD and AMD orderings. The mean-field Hamiltonian of the AMD can be explicitly written as:
    \begin{align} \label{eq:AMD matrix elements} \mathcal{\hat{H}}_{\rm AMD} = \frac{ \sqrt{3} h_{\rm AMD} }{10}    \left( \hat{c}^{\dagger}_{  z\uparrow}\hat{c}_{  x\downarrow} + \hat{c}^{\dagger}_{  z\downarrow} \hat{c}_{  x\uparrow}\right) + \text{h.c.}, \end{align} This, in accordance with $\lambda \hat{l}_y \hat{s}_y$ [Eq.~(\ref{eq:lysy matrix elements})], leads to the vertical splitting of a four-fold degeneracy at the $\Delta_1$ point at $E = -1$, where the spin and orbital components of the four bands undergo hybridization, as shown in Figs.~\ref{fig:Enlarged_view}(f) and \ref{fig:Enlarged_view}(h). Indeed, the energy splitting from $E=-1$ at the $\Delta_1$ point is calculated as follows:
    \begin{align} \label{eq:energy shift at Delta1} E(\Delta_1) \approx -1 \pm \left(\frac{-\sqrt{3}}{10} h_{\rm AMD} \pm \frac{1}{2} \lambda \right). \end{align} When $\lambda \neq 0$ and $h_{\rm AMD} = 0$, the Berry curvatures are generated but cancel out because the signs of the matrix elements of $\lambda \hat{l}_y \hat{s}_y$ between $\hat{c}^\dagger_{  z\downarrow} \hat{c}_{  x\uparrow}$ and $\hat{c}^\dagger_{  z\uparrow} \hat{c}_{  x\downarrow}$ are opposite due to the presence of the time-reversal symmetry, as shown in Eq.~(\ref{eq:lysy matrix elements}). However, when $\lambda \neq 0$ and $h_{\rm AMD} \neq 0$, the sign of the Berry curvature is modulated by the AMD, resulting in a net Berry curvature, as shown in Figs.~\ref{fig:Enlarged_view}(b) and \ref{fig:Enlarged_view}(d). This arises from the difference in sign of the matrix elements of $\mathcal{\hat{H}}_{\rm AMD}$ between $\hat{c}^\dagger_{  z\downarrow} \hat{c}_{  x\uparrow}$ and $\hat{c}^\dagger_{  z\uparrow} \hat{c}_{  x\downarrow}$ in Eq.~(\ref{eq:AMD matrix elements}).

    Similarly to the $\Delta_1$ point, the Berry curvature is also enhanced at the anticrossing band around $E \sim 0.8$ in the low-symmetry regions in the Brillouin zone, located along the $\Gamma$-M line, as shown in Fig.~\ref{fig:Hall_conductivity_and_its_source}(f). Meanwhile, the Berry curvature tends to be canceled out in the vicinity of the M point around $E = -0.40$ and the $\Gamma$ point around $E = 1.59$, since the spin-degenerate pairs at these points have the opposite signs of Berry curvature, as shown in Figs.~\ref{fig:Hall_conductivity_and_its_source}(d) and ~\ref{fig:Hall_conductivity_and_its_source}(f).

    \begin{figure*}[t]
        \centering \includegraphics[width=\textwidth]{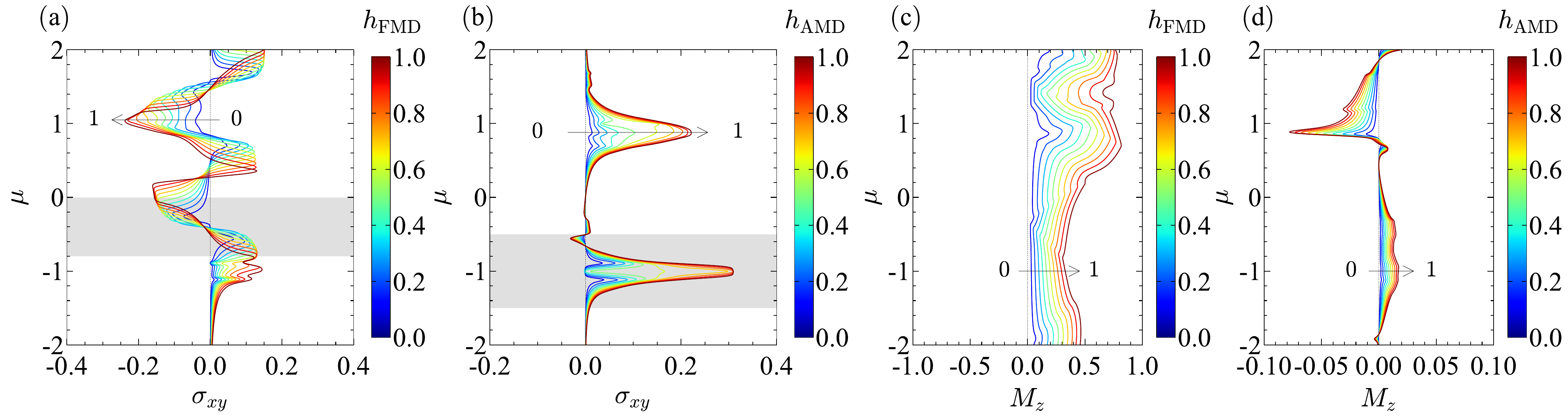}     
        \caption{
            (a, b) Anomalous Hall conductivity $\sigma_{xy}$ and (c, d) total magnetization $M_z$ as functions of the chemical potential $\mu$.
            The mean field of the FMD is varied within $0<h_{\rm FMD}<1$ for (a, c), and is fixed at $h_{\rm FMD}=0$ for (b, d).
            The mean field of the AMD is varied within $0<h_{\rm AMD}<1$ for (b, d), and is fixed at $h_{\rm AMD}=0$ for (a, c).
            The SOC constant is set to $\lambda=0.2$ for (a--d).
            The behaviors of $\sigma_{xy}$ in the highlighted regions of panels (a) and (b) are analyzed in detail in Figs.~\ref{fig:h_dependence_in_detail_sz} and \ref{fig:h_dependence_in_detail_Qzxsx}, respectively.
        }\label{fig:Overlay_mu_dependence}
    \end{figure*}    

    \begin{figure*}[t]
        \centering 
        \includegraphics[width=\textwidth]{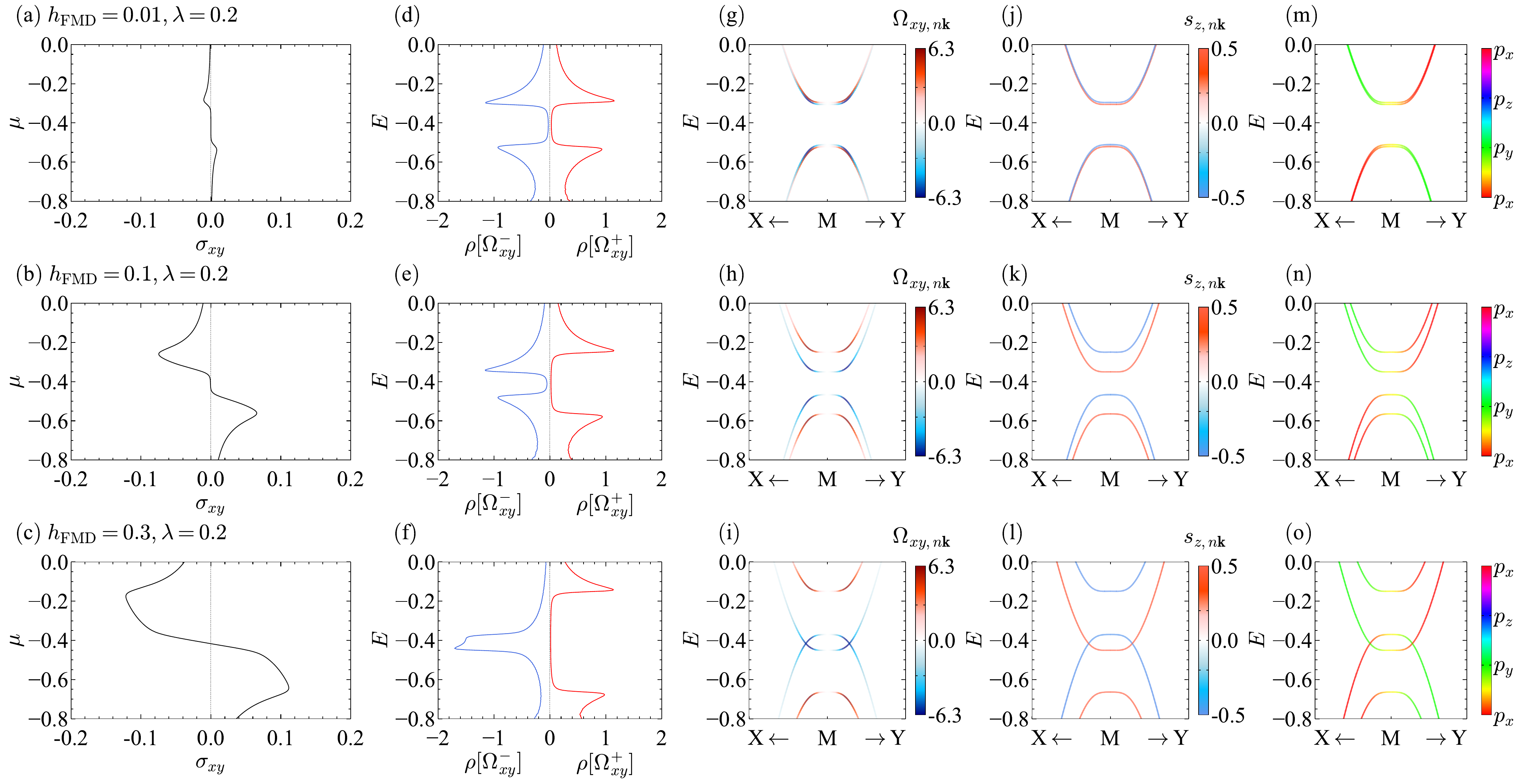}       
        \caption{
            (a, b, c) The anomalous Hall conductivity $\sigma_{xy}$ as a function of the chemical potential $\mu$.                                                                                                                   
            (d, e, f) Density of Berry curvature $\rho[\Omega_{xy}^{\pm}]$ as a function of the energy $E$.
            (g--o) Energy bands around the M point mapped with the following quantities: 
            (g, h, i) the Berry curvature $\Omega_{xy,n\bm{k}}$, 
            (j, k, l) the $z$-component of spin $s_{z,n\bm{k}} \equiv \langle \psi_{n\bm{k}} | \hat{s}_z | \psi_{n\bm{k}} \rangle$, and 
            (m, n, o) the orbital component $\sum_{\sigma}|\langle p_{\alpha\sigma} | \psi_{n\bm{k}}\rangle|^{2}$.
            The SOC constant is fixed at $\lambda = 0.2$ for (a--o). 
            The mean field of the FMD is set to $h_{\rm FMD} = 0.01$ for (a, d, g, j, m), $h_{\rm FMD} = 0.1$ for (b, e, h, k, n), and $h_{\rm FMD} = 0.3$ for (c, f, i, l, o). 
            The mean field of the AMD is set to $h_{\rm AMD} = 0$ for (a--o).
        }\label{fig:h_dependence_in_detail_sz}
    \end{figure*}
    
    \begin{figure*}[t]
        \centering 
        \includegraphics[width=\textwidth]{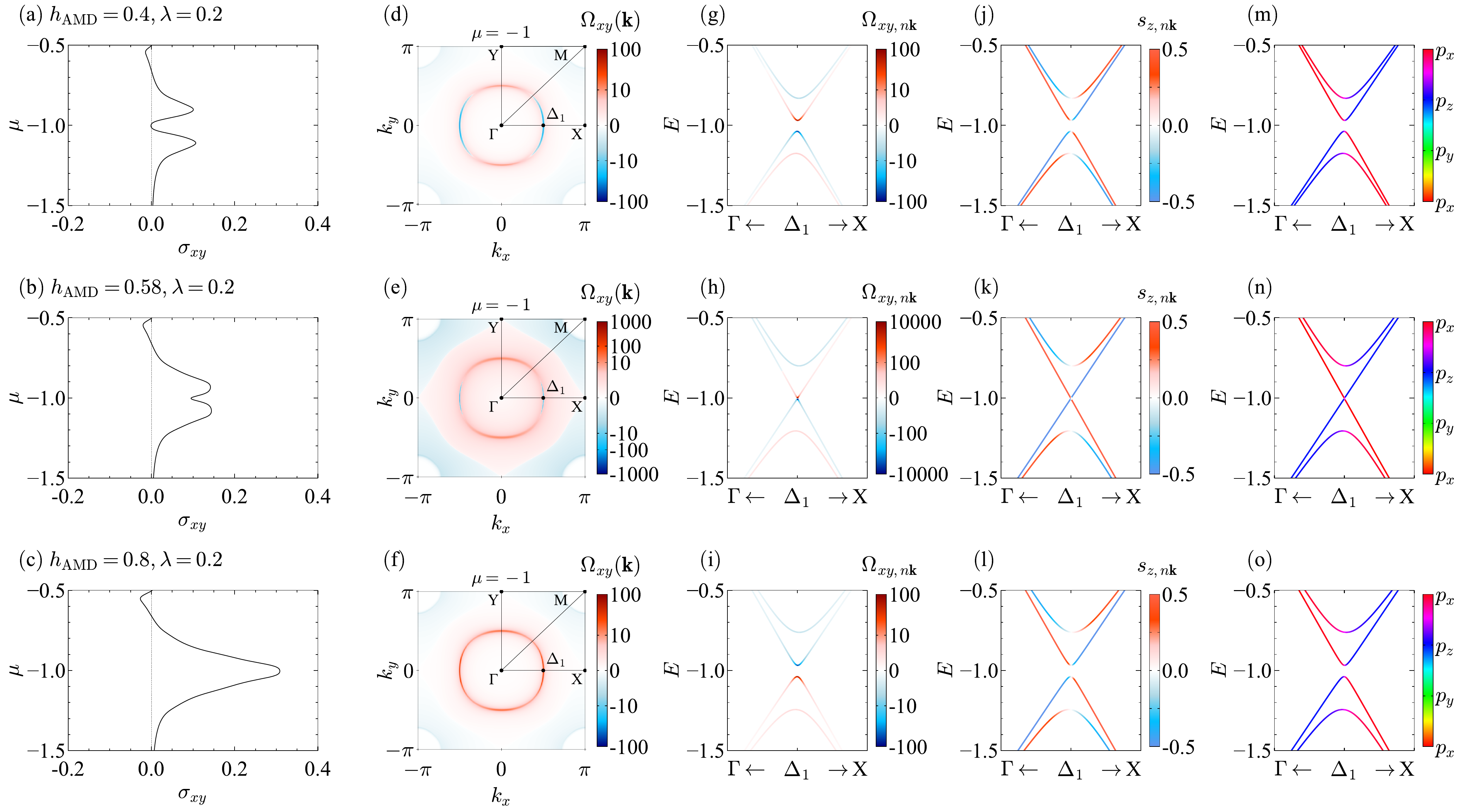}       
        \caption{
            (a, b, c) The anomalous Hall conductivity $\sigma_{xy}$ as a function of the chemical potential $\mu$.
            (d, e, f) Momemtum-resolved Berry curvature $\Omega_{xy}(\bm{k})$ at $\mu = -1$, mapped in the Brillouin zone.
            (g--o) Energy bands around the $\Delta_1$ point mapped with the following quantities: 
            (g, h, i) the Berry curvature $\Omega_{xy,n\bm{k}}$, 
            (j, k, l) the $z$-component of spin $s_{z,n\bm{k}} \equiv \langle \psi_{n\bm{k}} | \hat{s}_z | \psi_{n\bm{k}} \rangle$, and 
            (m, n, o) the orbital component $\sum_{\sigma}|\langle p_{\alpha\sigma} | \psi_{n\bm{k}}\rangle|^{2}$.                        
            The SOC constant is fixed at $\lambda = 0.2$ for (a--o). 
            The mean field of the FMD is set to $h_{\rm FMD} = 0$ for (a--o).
            The mean field of the AMD is set to $h_{\rm AMD} = 0.4$ for (a, d, g, j, m), $h_{\rm AMD} = 0.58$ for (b, e, h, k, n), and $h_{\rm AMD} = 0.8$ for (c, f, i, l, o). 
        }\label{fig:h_dependence_in_detail_Qzxsx}
    \end{figure*}
    
    \begin{figure*}[t]
        \centering \includegraphics[width=\textwidth]{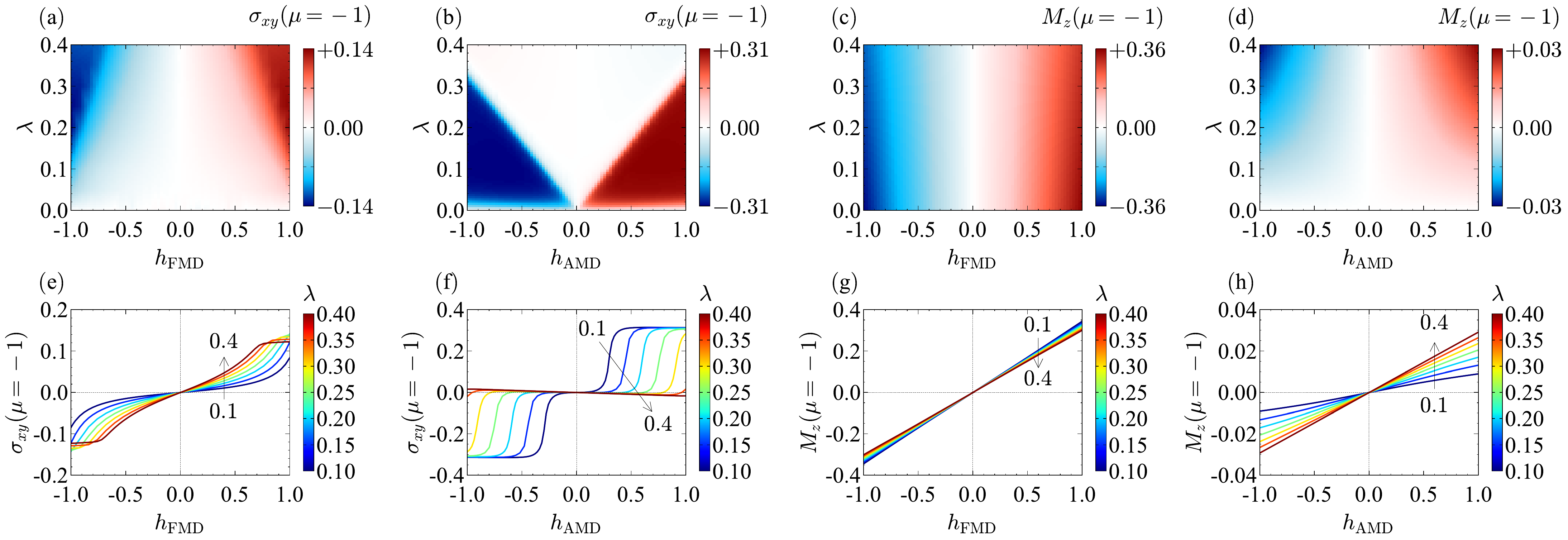}     
        \caption{
            (a, b) $\sigma_{xy}$ and (c, d) $M_z$, calculated as functions of the SOC constant and the magnetic mean field.
            (e, f) $\sigma_{xy}$ and (g, h) $M_z$ as functions of the magnetic mean field.
            (a, c, e, g) are calculated under the FMD ordering, and (b, d, f, h) under the AMD ordering.
            The SOC constant $\lambda$ varies from 0.1 to 0.4 in steps of 0.05 in (e--h).
            The chemical potential is fixed at $\mu = -1$ in (a--h).
        }\label{fig:heatmap-of-h-lam-dependence}
    \end{figure*}

\subsection{Parameter dependence of the AHE and magnetization}\label{subsec:Parameter_dependence}
    Next, we investigate the model parameter dependence of $\sigma_{xy}$ and $M_z$ with an emphasis on the SOC and magnetic mean field. Figure~\ref{fig:Mz_sigma_plot} shows the results illustrating the distinct trends between the FMD and AMD orderings with respect to $|M_z|$ and $|\sigma_{xy}|$. The data is plotted for different values of $\mu$, ranging from $-2$ to $2$. The overall results clearly indicate that the magnitude of $\sigma_{xy}$ under the AMD ordering is comparable to that under the FMD ordering, with the much smaller magnetization in the former compared to that in the latter. This means that large magnetization is not necessary for the realization of a large AHE.

    In addition, distinct trends are observed with changes in $\lambda$. For the weak SOC regime ($\lambda=0.05$) in Fig.~\ref{fig:Mz_sigma_plot}(a), the maximum values of $|\sigma_{xy}|$ under the AMD ordering exceed those under the FMD ordering for all the magnitudes of the magnetic mean field (0.2, 0.5, and 0.8). On the other hand, for the relatively large SOC regime ($\lambda=0.2$) in Fig.~\ref{fig:Mz_sigma_plot}(b), $|\sigma_{xy}|$ under the AMD tends to be suppressed with its maximum values being smaller than that under the FMD until the magnetic mean field reaches 0.8. Besides, the magnetization under the AMD ordering slightly increases with a larger SOC. This result implies that a small SOC, relative to the magnetic mean field, is favorable for achieving a larger $|\sigma_{xy}|$ with a smaller $|M_z|$, under the AMD ordering.

    Figures~\ref{fig:Overlay_mu_dependence}(a) and \ref{fig:Overlay_mu_dependence}(c) show $\sigma_{xy}$ and $M_z$ as functions of $\mu$, calculated for varying values of $h_{\rm FMD}$ under the FMD ordering. Both $|\sigma_{xy}|$ and $M_z$ monotonically increase with $h_{\rm FMD}$, and the peak positions of $\sigma_{xy}$ gradually shift. As illustrated in Fig.~\ref{fig:h_dependence_in_detail_sz}, this behavior stems from the spin splitting of the energy band near the high-symmetry point. Figure~\ref{fig:h_dependence_in_detail_sz}(g) shows the energy band at around the M point calculated under $h_{\rm FMD} = 0.01$, with the color indicating the Berry curvature. Since the magnetic mean field is small, the spin-up and spin-down bands are nearly degenerate [Fig.~\ref{fig:h_dependence_in_detail_sz}(j)]. Meanwhile, the orbital degeneracy of the $p_{y\uparrow\downarrow}$ and $p_{x\uparrow\downarrow}$ is split by $\pm 0.1$ due to the SOC with a coupling constant of $\lambda = 0.2$ [Fig.~\ref{fig:h_dependence_in_detail_sz}(m)]. The density of Berry curvature is finite under the SOC, although it almost cancels out, as shown in Fig.~\ref{fig:h_dependence_in_detail_sz}(d). This cancellation occurs because the SOC induces the opposite signs of Berry curvature [Fig.~\ref{fig:h_dependence_in_detail_sz}(g)] in the spin $\uparrow\downarrow$ bands, and their energy levels are nearly equal due to the small magnetic mean field [Fig.~\ref{fig:h_dependence_in_detail_sz}(j)]. Consequently, $\sigma_{xy}$ nearly vanishes over $-0.8 \leq \mu \leq 0.0$, as shown in Fig.~\ref{fig:h_dependence_in_detail_sz}(a).

    Then, as $h_{\rm FMD}$ increases from $0.01$ to $0.3$, the band splitting between the spin $\uparrow\downarrow$ states becomes more pronounced [Figs.~\ref{fig:h_dependence_in_detail_sz}(j, k, l)]. Correspondingly, the densities of positive and negative Berry curvature are gradually separated [Figs.~\ref{fig:h_dependence_in_detail_sz}(d, e, f)], resulting in an increase and a peak shift of $\sigma_{xy}$ [Figs.~\ref{fig:h_dependence_in_detail_sz}(a, b, c)]. It is noted that the magnitude of momentum-resolved Berry curvature $\Omega_{xy,n\bm{k}}$ remains unchanged for varying values of $h_{\rm FMD}$[Figs.~\ref{fig:h_dependence_in_detail_sz}(g, h, i)]. This is because the energy difference between the spin $\uparrow\uparrow$ or $\downarrow\downarrow$ bands remains unchanged [Figs.~\ref{fig:h_dependence_in_detail_sz}(j, k, l)], and the orbital components of the relevant bands remain the same [Figs.~\ref{fig:h_dependence_in_detail_sz}(m, n, o)] while the magnetic mean field $h_{\rm FMD}$ increases. Therefore, large $\sigma_{xy}$ is achieved when the bands with large Berry curvature are well separated, circumventing the cancellation between positive and negative contributions. Accordingly, $\sigma_{xy}$ saturates when the spin splitting becomes sufficiently large, minimizing the cancellation of Berry curvature.

    On the other hand, $\sigma_{xy}$ under the AMD ordering shows a step-function-like behavior with respect to the change in $h_{\rm AMD}$. Figures~\ref{fig:Overlay_mu_dependence}(b) and \ref{fig:Overlay_mu_dependence}(d) illustrate $\sigma_{xy}$ and $M_z$ as functions of $\mu$, respectively, for varying values of $h_{\rm AMD}$. When $\mu = -1$, $\sigma_{xy}$ remains nearly zero for $h_{\rm AMD} \lesssim 0.6$, increases sharply around $h_{\rm AMD} \approx 0.6$, and stays approximately constant at 0.3 for $h_{\rm AMD} \gtrsim 0.6$. This abrupt change in $\sigma_{xy}$ is presumably caused by a competition between the magnetic mean field of the AMD and the $y$ component of the SOC, which can be explicitly expressed as 
    \begin{align} \label{eq:AMD+SOC matrix elements} \mathcal{\hat{H}}_{\rm AMD} + \lambda \hat{l}_y \hat{s}_y &= \left(-\frac{\sqrt{3}}{10} h_{\rm AMD} + \frac{\lambda}{2} \right) \hat{c}^{\dagger}_{z\uparrow} \hat{c}_{x\downarrow} \notag \\ &+ \left(-\frac{\sqrt{3}}{10} h_{\rm AMD} - \frac{\lambda}{2} \right) \hat{c}^{\dagger}_{z\downarrow} \hat{c}_{x\uparrow} + \text{h.c.}, \end{align} where the sign change in the first term occurs at $h_{\rm AMD}=5\lambda / \sqrt{3} \approx 2.9\lambda$. This affects the sign of Berry curvature at $\Delta_1$ to reverse, resulting in the cancellation of Berry curvature for small $h_{\rm AMD}$. This intuitive interpretation is confirmed in Fig.~\ref{fig:h_dependence_in_detail_Qzxsx}, which presents the cases of $h_{\rm AMD} = 0.4, 0.58,$ and $0.8$ as examples to demonstrate the sign change in the Berry curvature and its impact on $\sigma_{xy}$. When $h_{\rm AMD} = 0.4$ and $\mu=-1$, the $p_x$-$p_z$ crossing in the $|k_y|<|k_x|$ region leads to negative Berry curvature[Fig.~\ref{fig:h_dependence_in_detail_Qzxsx}(g, m)], while the $p_y$-$p_z$ crossing in the $|k_y|>|k_x|$ region leads to positive Berry curvature[Fig.~\ref{fig:h_dependence_in_detail_Qzxsx}(d)]. As a result, Berry curvature between these two regions nearly cancels out in the entire Brillouin zone, resulting in a negligibly small $\sigma_{xy}$ at $\mu=-1$[Fig.~\ref{fig:h_dependence_in_detail_Qzxsx}(a)].

    At $h_{\rm AMD} = 0.58 \approx 5\times0.2/\sqrt{3}$ in Eq.~(\ref{eq:energy shift at Delta1}), one of the gap at $\Delta_1$ nearly closes; the magnitude of the Berry curvature reaches approximately $10^4$ at its maximum [Fig.~\ref{fig:h_dependence_in_detail_Qzxsx}(h)]. Additionally, the region in the Brillouin zone with negative Berry curvature shrinks [Fig.~\ref{fig:h_dependence_in_detail_Qzxsx}(e)], resulting in a finite value of $\sigma_{xy}$ at $\mu = -1$ [Fig.~\ref{fig:h_dependence_in_detail_Qzxsx}(b)]. Then, as $h_{\rm AMD}$ increases to $0.8$, Berry curvature in both $|k_y| < |k_x|$ and $|k_y| > |k_x|$ regions becomes positive [Fig.~\ref{fig:h_dependence_in_detail_Qzxsx}(f)] resulting in a large $\sigma_{xy}$ at $\mu=-1$ [Fig.~\ref{fig:h_dependence_in_detail_Qzxsx}(c)]. Note that during the change in the Berry curvature, the spin components [Fig.~\ref{fig:h_dependence_in_detail_Qzxsx}(j, k, l)] and orbital components [Fig.~\ref{fig:h_dependence_in_detail_Qzxsx}(m, n, o)] of the relevant bands remain nearly unchanged.

    As for $M_z$, its magnitude increases monotonically with the magnetic mean field in both the FMD and AMD orderings [Figs.~\ref{fig:Overlay_mu_dependence}(c, d)]. Nevertheless, the magnitude of $M_z$ in the AMD-ordered system is at least ten times smaller than that in the FMD-ordered system. In contrast to $\sigma_{xy}$, a step-function-like behavior is not observed in the magnetization under the AMD ordering.

    Figure~\ref{fig:heatmap-of-h-lam-dependence} shows $\sigma_{xy}$ and $M_z$ as a function of $(\lambda, h_{\rm FMD})$ or $h_{\rm AMD}$ at $\mu=-1$. In the FMD-ordered system, $\sigma_{xy}$ increases monotonically with respect to both $h_{\rm FMD}$ and $\lambda$, as shown in Figs.~\ref{fig:heatmap-of-h-lam-dependence}(a) and \ref{fig:heatmap-of-h-lam-dependence}(e). Meanwhile, $M_z$ increases monotonically with $h_{\rm FMD}$ but is almost independent of $\lambda$, as shown in Figs.~\ref{fig:heatmap-of-h-lam-dependence}(c) and \ref{fig:heatmap-of-h-lam-dependence}(g).

    In contrast, in the AMD-ordered system, $\sigma_{xy}$ exhibits an abrupt change at $|h_{\rm AMD}| = 5\lambda/\sqrt{3}$, below which $\sigma_{xy}$ is canceled to zero, as shown in Figs.~\ref{fig:heatmap-of-h-lam-dependence}(b) and \ref{fig:heatmap-of-h-lam-dependence}(f), which presumably results from the competition between $\hat{l}_y \hat{s}_y$ and $\hat{M}_{z}'^{(s_x)}$, as discussed earlier in this section. For large $h_{\rm AMD}$, $\sigma_{xy}$ takes constant values irrespective of $\lambda$.

    Finally, let us show the important model parameters to induce the magnetization under the FMD and AMD orderings by expanding the quantity $M_z = \operatorname{Tr}[e^{-\beta \mathcal{\hat{H}}} \hat{M}_z]$ in terms of $\beta \mathcal{\hat{H}}$~\cite{Hayami_PhysRevB.102.144441, Oiwa_doi:10.7566/JPSJ.91.014701}. The same technique has recently been used to obtain the essential model parameters for one-body physical quantities~\cite{Hayami_PhysRevB.101.220403} and nonlinear conductivity tensors~\cite{Yatsushiro_PhysRevB.105.155157, Hayami_PhysRevB.106.024405}. In the FMD-ordered system, one obtains 
    \begin{align} \label{eq:Mz_under_FMD} M_{z,\rm FMD} \propto h_{\rm FMD} .\end{align} Meanwhile, in the AMD-ordered system, $M_z$ shows a distinct model parameter dependence as:
    \begin{align} \label{eq:Mz_under_AMD} M_{z,\rm AMD} \propto h_{\rm AMD} \lambda_x, \end{align} where $\lambda_x$ represents the coefficient for the $x$ component of the SOC, i.e., we temporarily split the SOC Hamiltonian into $\sum_{\alpha=x,y,z} \lambda_{\alpha} \hat{l}_{\alpha} \hat{s}_{\alpha}$. This is consistent with the results shown in Figs.~\ref{fig:heatmap-of-h-lam-dependence}(d) and \ref{fig:heatmap-of-h-lam-dependence}(h), where $M_z$ increases with increasing $h_{\rm AMD}$ and $\lambda$. Therefore, it is confirmed that the SOC is essential for inducing magnetization under the AMD ordering.

\section{Summary}\label{sec:summary}
    We theoretically investigated the AHE under the FMD and AMD orderings by focusing on their similarity and difference from the microscopic viewpoint. By examining the momentum-resolved Berry curvature in each case, we found that the dominant contributions to the AHE are different, even though the magnitudes of Berry curvature are similar. In the FMD-ordered system, the AHE is dominated by the Berry curvature around the high-symmetry points, such as the M and $\Gamma$ points, where the spin splittings by the FMD order parameter give rise to large AHE. On the other hand, in the AMD-ordered system, $\sigma_{xy}$ is enhanced when the band anticrossings occur at the low-symmetry points. Accordingly, the chemical potential dependence of the anomalous Hall conductivity is different from each other. In addition, we found that the total magnetization under the AMD ordering is much smaller than that under the FMD ordering.

    We also observed that in the FMD-ordered system, the AHE increases with a larger FMD mean field but eventually saturates, whose behavior is accounted for by the Zeeman-type spin splitting at high-symmetry points in the Brillouin zone. When The mean field of the FMD is small, the spin-degenerate bands with the opposite signs of Berry curvature overlap and cancel out. As the mean field of the FMD increases, the bands are well separated, avoiding the cancellation of the Berry curvature, which results in a large AHE. Meanwhile, the AHE saturates when the spin splitting is large enough since the magnitude of AHE does not depend on the magnitude of the mean field of the FMD. Therefore, the lifting of the spin-degenerated bands is essential for inducing the AHE under the FMD ordering. In other words, the emergence of AHE is related to the uniform magnetization.

    In contrast, in the AMD-ordered system, the AHE shows a step-function-like behavior with the changes in the magnetic mean field, $h_{\rm AMD}$. This behavior is attributed to the opposite signs of the matrix elements of the SOC and AMD mean field. As a result, the Berry curvature at the band anticrossings changes its sign in part of the Brillouin zone, depending on the relative magnitude of the SOC and AMD mean field. Specifically, when the SOC becomes smaller than the mean field of the AMD, the AHE is enhanced, which can be larger than that under the FMD ordering. Thus, one can expect a large AHE without the total magnetization under the AMD ordering.

    The AMD moment is induced by both collinear and noncollinear AFMs, provided that the magnetic symmetry of the AFM structures aligns with that of the FM structure. Therefore, our findings will provide valuable insights for future experiments in a wide range of AFM materials, offering guidelines on controlling the anomalous Hall conductivity in various magnetic materials, and paving the way for new applications in AFM spintronics.

\begin{acknowledgments} This research was supported by JSPS KAKENHI Grants Numbers JP21H01037, JP22H00101, JP22H01183, JP23H04869, JP23K03288, JP23K20827, and by JST CREST (JPMJCR23O4) and JST FOREST (JPMJFR2366).   Parts of the numerical calculations were performed in the supercomputing systems in ISSP, the University of Tokyo.
\end{acknowledgments}

\bibliographystyle{apsrev4-2}
\bibliography{ahe}

\end{document}